\documentclass[aps,showpacs,preprint]{revtex4}

\usepackage{amsmath}
\usepackage{amsfonts}
\usepackage{bm}
\usepackage{graphicx}
\usepackage{color}

\begin{document}
\title{Quantum scaling in  nano-transistors}
\author{U. Wulf}
\affiliation{Technische Universit\"at Cottbus, Fakult\"at 1, Postfach 101344,
             03013 Cottbus, Germany }

\begin{abstract}
In our previous papers on ballistic quantum transport in nano-transistors
[J. Appl. Phys. 98, 84308 (2005)] it was demonstrated
that  under certain conditions it is possible to reduce the three-dimensional transport problem
to an effectively one-dimensional one. 
We show that such an effectively one-dimensional description can be
cast in a scale-invariant form.  We obtain dimensionless variables
for the characteristic channel length $l$ and width of the transistor which
determine the scale-invariant output characteristic. 
For $l \gtrsim 10$, in the strong barrier regime, the output
characteristics are similar to that of a conventional MOSFET assuming
an ideal form for $l \rightarrow \infty$. In the weak barrier regime,  $l \lesssim 10$,
strong source-drain currents lead to
i-v characteristics that differ qualitatively
from that of a conventional transistor.
Comparing with experimental data we find qualitative agreement.
\end{abstract}


\maketitle

\begin{figure} [t]
\includegraphics[width=7cm]{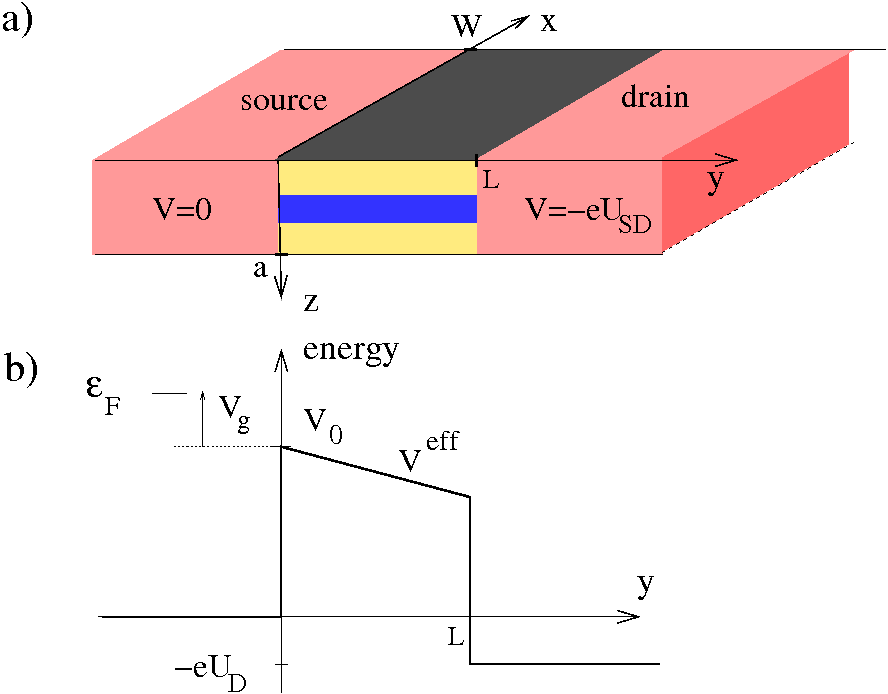}

\caption{
a) Generic n-channel FET  in three dimensions: narrow electron channel (blue)
with abrupt transition to source and drain contact (red).
b.) Effective one-dimensional potential in the
ON-state: $V_g \equiv \epsilon_F - V_0 >0$.
}

\label{1dreimod}
\end{figure}

Recently, a number of  extremely small nano-transistors with physical gate lengths
of fifteen nanometers or less have been fabricated\cite{nec,intel,amd}.
A rather simple but particularly illuminating approach to describe the electron dynamics in such small
structures is to assume ballistic quantum transport (for a recent review see Ref. 
[\onlinecite{lundstrom06}]). It is found
for a long enough electron channel that the output characteristic is
similar to that of a conventional transistor.
However, in this quantum ballistic transistor (QBT) regime  
source-drain tunneling currents cause
residual slopes in the ON-state 
and the development of only a quasi-OFF-state. 
In Refs.\ [\onlinecite{nemnes04,nemnes05}] it was shown that
separating the ON-state 
of classically allowed transport and the quasi OFF-state of tunneling transport
there is a threshold characteristic (TH) which exhibits a close-to-linear
dependence of the current on the drain voltage. Above the TH,
in the ON-state, the I-V curves are characterized by a negative bending
and below the TH by a positive bending.
In Refs.\ [\onlinecite{nemnes04,nemnes05}]  it was
furthermore demonstrated that for a narrow electron
channel the complete three-dimensional quantum problem
can be approximated by an effectively one-dimensional one 
(Fig.\ \ref{1dreimod}).

In the present paper we recast our effectively one-dimensional description 
of quantum transport in a nano-transistor in a scale-invariant form. 
The scaled output characteristics of a transistor are  governed by its 
dimensionless characteristic length $l= L/L_0$ and the width $w = W/(\pi L_0)$,
which are given by its physical length $L$ and width $W$ 
in units of the scaling length $L_0 = \hbar (2 m^*  \epsilon_F)^{-1/2}$. Here
$\epsilon_F$ is the Fermi energy in the source contact 
and $m^*$ is the effective mass
in the electron channel in transport direction. 
The characteristic length permits a classification of
source-drain barriers in terms of their strength.
In the strong barrier regime, $l \gtrsim 10$, typical output
characteristics in the QBT regime occur. 
As $l \rightarrow \infty$, ideal output characteristics (IOC)
are assumed  with a vanishing residual slope in
the ON-state and a clear OFF-state
exhibiting negligible leakage currents.
 In the weak barrier regime, $l \lesssim 10$, 
strong source-drain currents lead to  I-V characteristics which differ strongly
from that of a conventional transistor.
In the limit of large $w$
we relate our scaling theory to the experiments and find qualitative agreement.

In Fig.\ \ref{1dreimod} our effective potential $V^{eff}$
is illustrated. Though simple, it allows us
to define the essential parameters: 
because of good screening it is constant
in the source contact with $V^{eff}(y \leq 0) =0$ and in the drain contact
with $V^{eff} (y \geq L) = -eU_{D}$, where $U_{D}$
is the applied drain voltage. The  source-drain barrier is taken
as a constant potential offset  $V_0$ for $0 \leq y  \leq L$. 
In the latter interval there is an additional contribution
due to the applied drain voltage which is assumed
to  fall off linearly so that $V^{eff} (0 \leq y \geq L) = V_0 -eU_{D} y/L$. 
Such a piecewise linear potential\cite{ueno02}
or more complex potentials with the same qualitative behavior\cite{lundstrom06}
have been used in the literature. We calculate the drain current through
\begin{equation}
i  =  \hat{\epsilon}_F^{-1} \int_{0}^{\infty} d\hat{\epsilon} \,
\left[
  F  ( \hat{\epsilon}_F - \hat{\epsilon}  )
- F  ( \hat{\epsilon}_F - \hat{v}_{D} - \hat{\epsilon} )
\right]
T^{eff}_{\beta \hat{v}_{D}} (\hat{\epsilon}),
\label{tsunorm}
\end{equation}
with  the normalized drain current $i = I/I_0$, $I_0 = 2 e \epsilon_F /h$,
and $\hat{\epsilon}_F = \epsilon_F /V_0$, and $\hat{v}_{D} = e U_{D}/V_0$.
The current transmission
$T^{eff}_{\beta \hat{v}_{D}} (\hat{\epsilon}) = 
k^{eff}_D(\hat{\epsilon}) |t^S (\hat{\epsilon})|^2  [k^{eff}_S(\hat{\epsilon}))]^{-1}$ 
results
from the transmission coefficient $t^S (\hat{\epsilon})$ of a source-incident scattering function
in a one-dimensional scattering problem associated with the
effective Schr\"odinger equation
\begin{equation}
\left[
- {1 \over \beta} {d^2 \over d \hat{y}^2} + \hat{v}^{eff}(\hat{y}) - \hat{\epsilon}
\right] \psi (\hat{y},\hat{\epsilon}) = 0.
\label{scalschroe}
\end{equation}
Here we have $\hat{y} = y/L$, $\beta = 2 m^* V_0 L^2 / \hbar^2$,
$\hat{v}^{eff}(\hat{y}) = V^{eff}(y)/V_0$,
$k_s^{eff} = \sqrt{ \beta  (\hat{\epsilon} + \hat{v}_s)}$, $s=S/D$, and
$v_S=0$.  The three-dimensional geometry of the transistor determines the
choice of the supply function $F$.
For a narrow transistor (small $W$) it
was shown in Ref.\ [\onlinecite{nemnes05}] that at $T=0$ it is given by
$F(x)=\Theta(x)$.
In a straightforward way one can generalize this result to a wide transistor 
yielding $ F(x) = \hat{w} \sqrt{x} \Theta(x)$,
with the normalized transistor width $\hat{w} =  \sqrt{ 2 m^* W^2 V_0} / ( \hbar \pi )$
\cite{wulfun}.
\begin{figure}[b]
   \begin{center}
   \noindent\includegraphics[width=6.5cm]{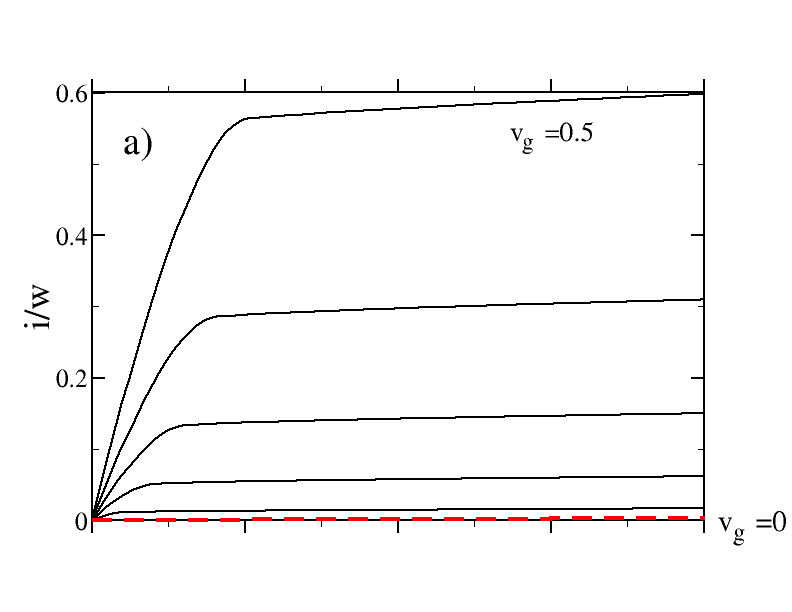}

   \vspace*{-.7cm}

   \noindent\includegraphics[width=6.5cm]{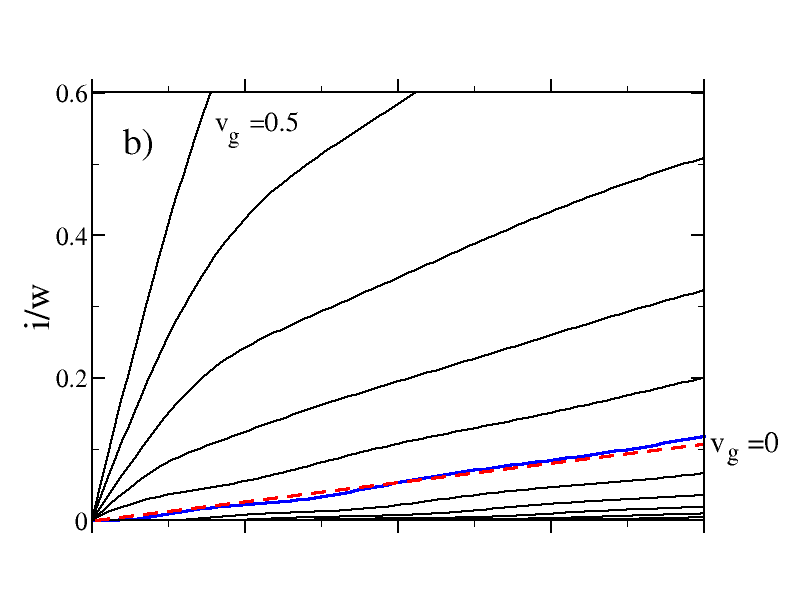}

   \vspace*{-.7cm}

   \noindent\includegraphics[width=6.5cm]{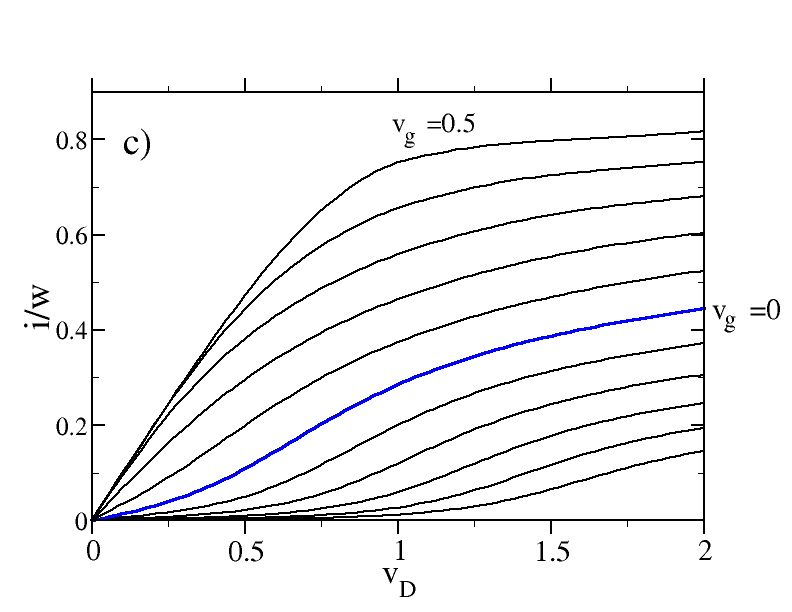}
\end{center}
\caption{i-v-traces in wide transistor limit, 
$v_g$ starting from 0.5 with decrements of 0.1 (solid lines, blue for
$v_g=0$).  a) $l=500$
and  b) $l=10$ in strong barrier regime. 
Best straight line fit for $v_g=0$ in red dashed line, 
in a) coinciding with the x-axis.  c) $l=5$ in weak barrier regime.
}
\label{2iv}
\end{figure}

To consider an I-V chart $I (U_{G},U_{S})$ we represent the gate potential $U_{G}$
by the parameter $V_g = \epsilon_{F} - V_0$, which is the deviation
of the Fermi energy in the source contact from the maximum of the source-drain barrier 
(see Fig.\ \ref{1dreimod} (b)).
Normalized to $V_0$ one has $\hat{\epsilon}_F =  \hat{v}_g + 1$,
allowing us to eliminate $\hat{\epsilon}_F$ in Eq.\ (\ref{tsunorm}). 
Furthermore, we define new variables $v_{D} = e U_{D} / \epsilon_F$
and $v_g = V_g/\epsilon_F$. These
are normalized to $\epsilon_F$, which is
independent of the gate voltage.
To cast $\hat{v}_{D}$ and $\hat{v}_g$ in terms of $v_{D}$ and $v_g$
in Eq.\ (\ref{tsunorm}) one exploits the identities
$\hat{v}_g = v_g /(1 - v_g)$ and
$\hat{v}_{D} = v_{D} /(1 - v_g)$.
Furthermore, the parameter $\beta$ 
is replaced by $\beta = l^2 (1 - v_g)$,
where we introduce the dimensionless characteristic length of the transistor as
$l = L/L_0$  with the scaling length
\begin{equation}
L_0 =  {\hbar \over \sqrt{2 m^*  \epsilon_F }}.
\label{lo}
\end{equation}  
Likewise $\hat{w} = w  (1 - v_g)$,
where we introduce the characteristic width of the transistor as
$w =  W/ (\pi L_0)$.

\begin{figure}
\begin{center}
 \includegraphics[width=6.6cm]{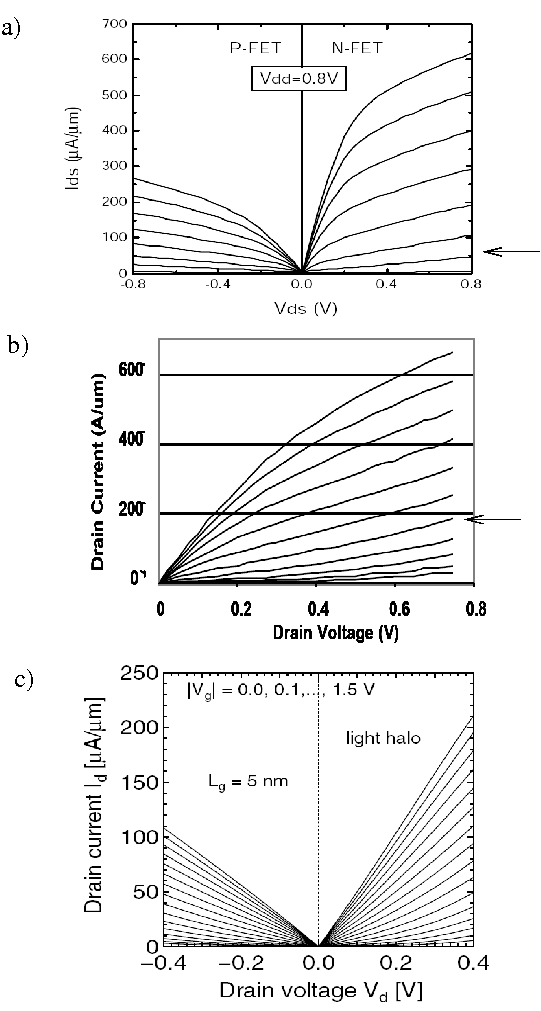}
\end{center}
\caption{Experimental output characteristics:
a) 15nm gate length  (taken from Ref. [\onlinecite{amd}])
b) 10nm gate length  (taken from Ref. [\onlinecite{intel}]),
We locate the
position of the TH in the strong barrier limit as marked by a blue arrow.
c) 5nm gate length (taken from Ref. [\onlinecite{nec}]).}
\label{3traexp}
\end{figure}

The calculated $I-V$ traces in Figs.\ \ref{2iv} (a) and (b) exemplify
the QBT regime at  $l \gtrsim 10$.
For positive $v_g$, i.e. in the ON-state,
an initial linear dependence of the drain current for small drain voltages
turns into a quasi-saturation regime for larger drain voltages. 
The turnover between linear and quasi-saturation regime becomes more and more abrupt,
which is seen in the experimental transistors in Fig.\ \ref{3traexp} (a) and (b) as
well.  The essentially linear THs at  $v_g=0$ 
allow us to define a threshold conductivity through $ i/w = \sigma^{th} v_D$.
Upon analysis of the numerical data it can be seen that in the range 
of interest $10 \leq l \leq 500$ one can approximate with an error of less than 10 percent
$\sigma^{th} \sim 0.94 l^{-1.25}$. This weak power law decrease indicates
that the IOC are assumed slowly with growing $l$
and that there is no critical characteristic length to define a hypothetical
`ideal transistor regime´. 
As shown in Fig.\ \ref{2iv} (c)
in the weak barrier limit, $l \lesssim 10$, 
I-V traces differ strongly from that of a conventional field effect
transistor because of strong source-drain tunneling 
and no TH is observed.

For a rough estimate of the characteristic lengths in the experimental nano-transistors 
in Fig.\ (\ref{3traexp})
we approximate $\epsilon_F$ by the well-known expression for  a three-dimensional
non-interacting electron gas. Assuming  a high level of source-doping of
$N_D \sim 5 \times 10^{20} cm^{-3}$ yields
$\epsilon_F = 0.2 eV$. Inserting in Eq.\ (\ref{lo})
furthermore the silicon light mass, $m^* = 0.19 m_0$,
one  obtains $L_0 \sim 1 nm$.
The experimental channel lengths of $5nm$ to $15nm$ then correspond to
characteristic lengths of five, ten, and   fifteen.
As expected we find for  $l \gtrsim 10$ a close-to linear TH with an
experimental threshold slope $\Sigma^{th} = I/ (W U_{D}) =75 \mu A /(\mu m A) $
for $L=15nm$ and $\Sigma^{th} = 270 \mu A /(\mu m A)$ for $L=10nm$.
Writing $\Sigma^{th} = 2 \sigma^{th} e^2/(h\pi L_0)$ one obtains 
for $l=15$ a theoretical value for $\Sigma^{th}$ of $ 750 \mu A /(\mu m A)$
and for $l=10$ a theoretical value of $1250 \mu A /(\mu m A)$.
While the decrease in the threshold conductivity results
theoretically and experimentally, there is a considerable quantitative 
discrepancy which is to be expected because in our model
essential effects like heating of the transistor and microscopic Coulomb
interaction are not included.
Consistent with  Fig.\ \ref{2iv} (c) 
the I-V characteristics of the experimental transistor with 5nm  gate length
in Fig.\ \ref{3traexp} (c)
seem to deviate considerably from the
output characteristic of  conventional transistor. However,
for a characterization of the breakdown of the QBT-regime for
weak barriers further research is needed.
Finally, we note that in the range of the experimental $L$ the
barrier strength could be increased using metal contacts
like in a Schottky barrier transistor\cite{knoch07}.
Here larger $\epsilon_F$ are possible which
according to Eq.\ (\ref{lo}) lead to smaller $L_0$ and larger $l$.

In summary, we present a scaling theory for quantum transport in  nano-transistors.
The scaled i-v characteristics depend on the dimensionless characteristic length
of the transistor channel which allows for
a classification  of a given source-drain barrier in terms of its strength.
In the strong barrier regime, $l \gtrsim 10$, the output
characteristics are similar to that of a conventional MOSFET assuming
an ideal form for $l \rightarrow \infty$. In the weak barrier regime strong
source-drain currents occur and the i-v characteristics differ strongly
from that of a conventional transistor.

We thank J. Emtage for valuable discussions.

\end{document}